\documentclass[a4paper,11pt]{article}
\usepackage{pos}
\usepackage{float}
\usepackage{wrapfig}
\usepackage{cleveref}
\usepackage{subcaption}
\usepackage{booktabs}
\usepackage{enumitem}
\usepackage{titlesec}
\usepackage[left]{lineno}
\title{Enhancements to the IceCube Extremely High Energy Neutrino Selection using Graph \& Transformer Based Neural Networks}
\ShortTitle{Enhancing the IceCube EHE Neutrino Selection}

\author{The IceCube Collaboration \\{\normalsize \normalfont(a complete list of authors can be found at the end of the proceedings)}\\}

\emailAdd{maxwell.nakos@icecube.wisc.edu}
\emailAdd{arosted@icecube.wisc.edu}
\emailAdd{lulu@icecube.wisc.edu}

\abstract{

KM3NeT has recently reported the detection of a very high-energy neutrino event, while IceCube has previously set upper limits on the differential neutrino flux above 100 PeV but has yet to observe a neutrino event with an energy comparable to that of the KM3NeT detection. To improve diffuse measurements above 10 PeV, we apply machine learning techniques to enhance atmospheric muon background rejection and directional reconstruction. We utilize a Graph Neural Network (GNN) to perform a classification task that distinguishes neutrinos from high-energy atmospheric muons. The method allows for the rejection of early hits from laterally spread, lower-energy muons in cosmic ray showers without relying on directional reconstruction as a prior. Additionally, a Transformer-based Neural Network is implemented for directional reconstruction. Unlike previous likelihood-based rapid reconstruction algorithms that assume a single muon track, this method makes no prior assumptions about event topology of the particle inside the detector. We demonstrate improved background rejection and reconstruction performance using machine learning techniques. Applications to the development of future Extremely High Energy (EHE) selections are also discussed.

\vspace{4mm}

{\bfseries Corresponding authors:}

Maxwell Nakos$^{1*}$, 
Aske Rosted$^{2}$, 
Lu Lu$^{1}$\\ 

{$^{1}$ \itshape University of Wisconsin-Madison}\\
{$^{2}$ \itshape Chiba University}\\[4mm]
$^*$ Presenter
}

\ConferenceLogo{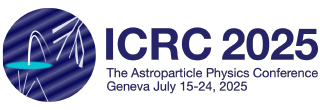}

\FullConference{39th International Cosmic Ray Conference (ICRC2025)\\
 15–24 July 2025\\
Geneva, Switzerland\\}
\renewcommand{\printHeadAuthors}{Maxwell Nakos, Aske Rosted, and Lu Lu}
\titlespacing*{\section} {1pt}{1ex}{1ex}
\titlespacing*{\subsection} {1pt}{1ex}{1ex}
\begin{document}
\maketitle

\section{Introduction}
Neutrinos with energies greater than 10 PeV offer valuable insight into the sources of ultra-high energy cosmic rays.
While magnetic fields deflect cosmic rays, neutrinos travel in straight lines, pointing directly back to their sources.
Additionally, due to the GZK effect, extragalactic cosmic rays ($E > 10^{19.5}$ GeV) could interact with cosmic microwave background, producing a cosmogenic neutrino flux~\cite{gzk}. The IceCube neutrino observatory is a cubic-kilometer neutrino detector situated at the geographic South Pole. The detector consists of an inice array of digital optical modules (DOMs) and a surface array (IceTop) for detecting cosmic rays~\cite{icecubedetector}. At the highest energies, Earth absorption~\cite{2017} significantly reduces the neutrino flux, making atmospheric muons the dominant background in IceCube's extremely-high-energy (EHE) neutrino searches. Previous selections have relied on overburden~\cite{bertandernie, 2013, 7year, 9year} or most recently, on stochastic energy loss profiles~\cite{EHE2025} to reject atmospheric muons. In this contribution, we investigate machine learning techniques to improve background rejection by classifying neutrinos against atmospheric muon bundles and directional reconstruction for the development of a future EHE neutrino selection.

\section{Atmospheric Muon Background Rejection}
\subsection{Lateral Distribution of Muon Background}
\begin{figure*}[ht]
    \centering
    \begin{subfigure}[t]{0.345\textwidth}
        \centering
        \includegraphics[width=1\textwidth]{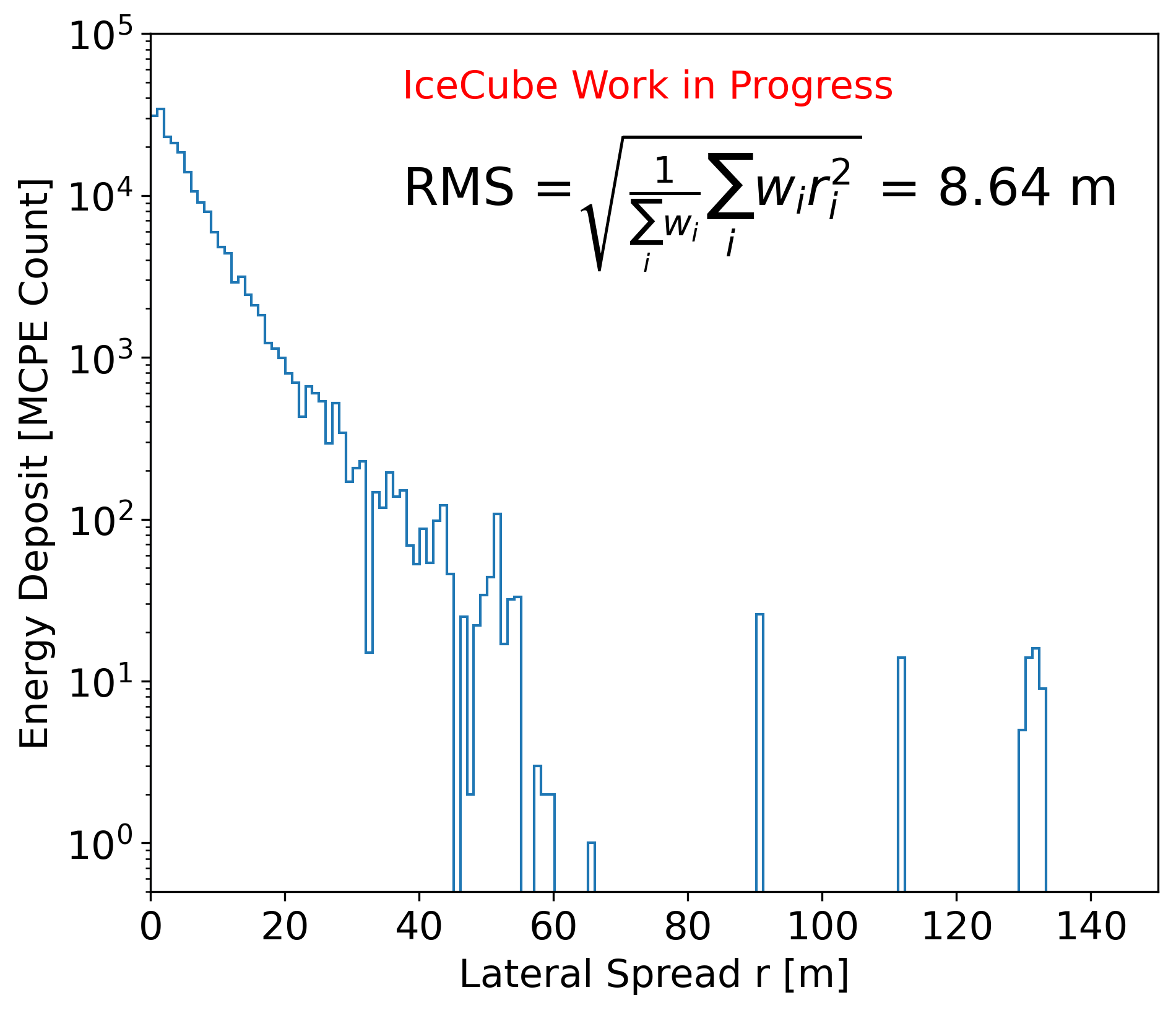}
        \captionsetup{skip=0pt}
        \caption{}
        \label{fig:rms_explanation}
    \end{subfigure}%
    ~
    \begin{subfigure}[t]{0.655\textwidth}
        \centering
        \includegraphics[width=1\textwidth]{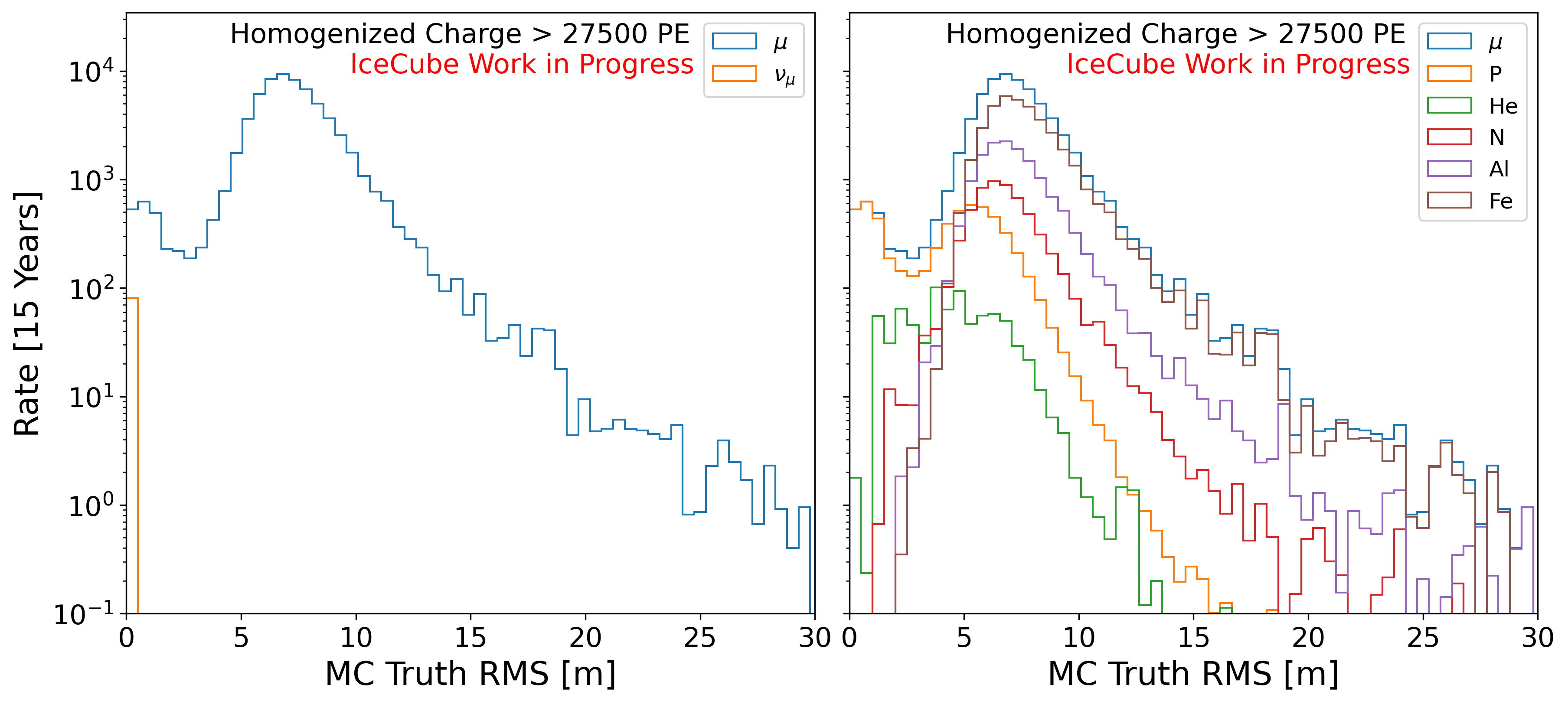}
        \captionsetup{skip=0pt}
        \caption{}
        \label{fig:numurms}
    \end{subfigure}%
    \captionsetup{skip=0pt}
    \captionsetup{belowskip=-10pt}
    \caption{\textbf{(a)} Lateral spread distribution function of a high-energy muon bundle. The RMS (root mean square) is computed by the equation shown in the plot, where $w_i$ is the number of photo electrons detected from an energy deposit and $r_{i}$ is the radial distance of the energy deposit from the highest energy muon in the event. \textbf{(b)} (left) RMS over an ensemble of cosmic ray showers and  $\nu _{\mu }$ events. The astrophysical spectrum is weighted to the best single power law measurement using starting tracks and the cosmic ray spectrum is weighted to H4a. (right) The RMS of cosmic rays broken down by the primary particle composition. The blue line shows the sum over all compositions.}
\end{figure*}

\vspace{-5pt}
The core challenge in EHE selections is to distinguish throughgoing high-energy neutrinos from cosmic ray showers, which produce large "bundles" of muons with lateral spread, whereas muon neutrino charge current (CC) interactions are likely to produce a high-energy muon (for $\nu_{\tau}^{CC}$, $\tau$) which deposits energy along a thin track through the detector. We seek to enhance rejection by incorporating lateral information in addition to the longitudinal energy loss profile. \begin{figure*}[ht]
    \centering
    \begin{subfigure}[t]{0.615\textwidth}
        \centering
        \includegraphics[width=1\textwidth]{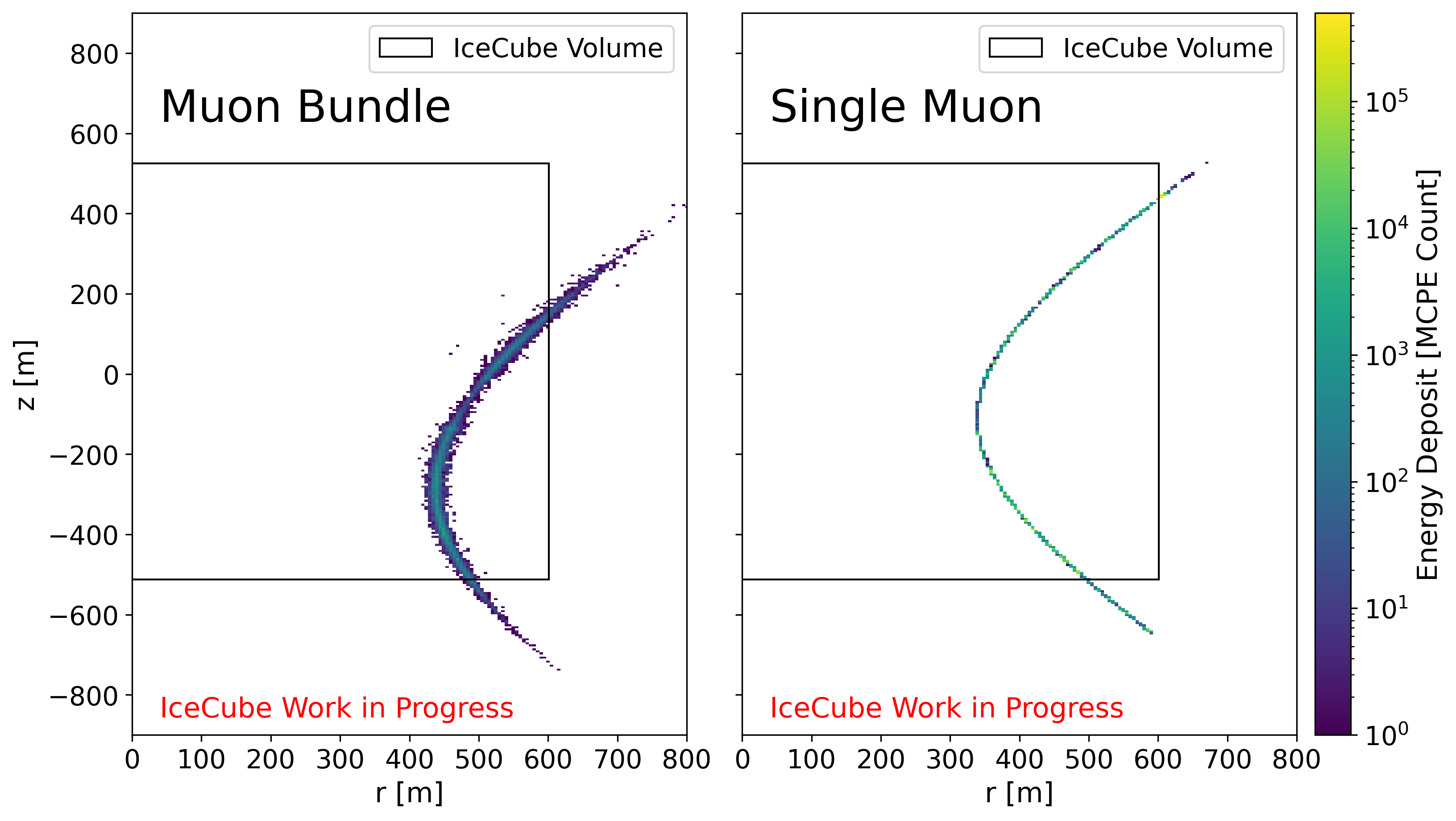}
        \captionsetup{skip=0pt}
        \caption{}
        \label{fig:singlebundle}
    \end{subfigure}%
    ~ 
    \begin{subfigure}[t]{0.385\textwidth}
        \centering
        \includegraphics[width=1\textwidth]{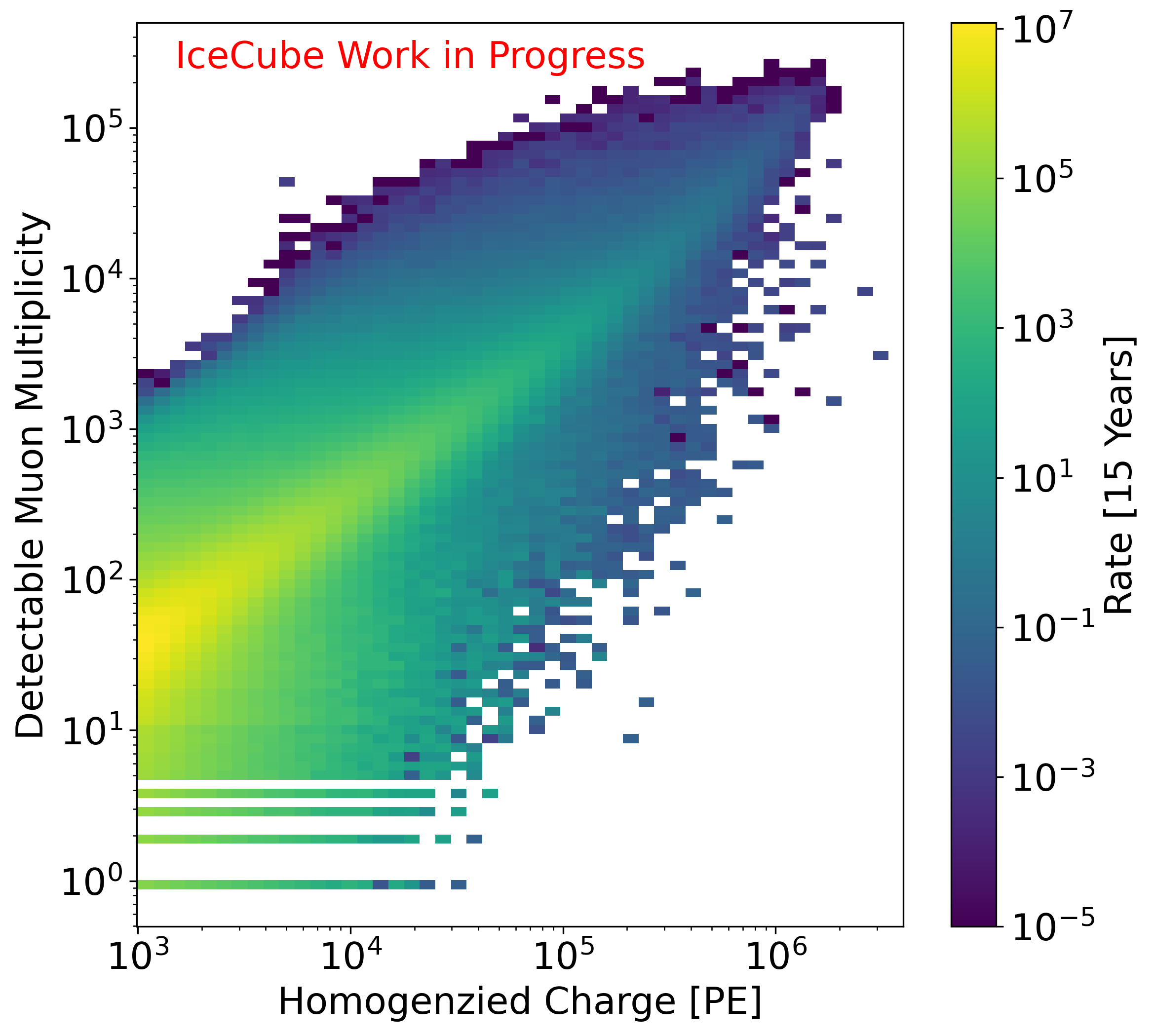}
        \captionsetup{skip=0pt}
        \caption{}
        \label{fig:multiplicity}
    \end{subfigure}%
    \captionsetup{skip=0pt}
    \captionsetup{belowskip=-10pt}
    \caption{\textbf{(a)} Energy deposit locations of events in IceCube, shown in cylindrical detector coordinates and weighted by the number of photoelectrons detected. Left: a cosmic ray shower with a bundle of muons exhibiting lateral spread from the trajectory of the initial cosmic ray. Right: a muon produced from a $\nu_{\mu}^{CC}$ interaction is shown traversing through the detector. \textbf{(b)} Number of muons in cosmic ray showers with a primary energy of greater than 1 PeV that register an in-ice pulse (excluding IceTop). Homogenized charge refers to the total photoelectrons (PE) in the detector excluding DeepCore and DOMs that collected more than half of the event charge.}
    \captionsetup{belowskip=0pt}
    \label{fig:multiplcitymetrics}
\end{figure*} \Cref{fig:singlebundle} shows the distribution of simulated photoelectron hits from the locations of the muon energy losses in the ice. The muon bundle creates wider deposition profile along its leading trajectory compared to a single muon track. 
\Cref{fig:multiplicity} shows the number of muons in a cosmic ray showers with primary energies above 1 PeV. As the charge threshold increases, cosmic ray showers are increasingly dominated by multiple muons. We set a charge threshold of 27,500 PE to filter out cosmic ray showers with single muons, based on \cref{fig:multiplicity}. \Cref{fig:rms_explanation} shows the root mean square (RMS) of the energy deposits (counted by MCPEs) from the leading trajectory for a single cosmic ray shower. In \cref{fig:numurms}, we see there is great separation power between neutrinos and cosmic rays in the lateral spread. Heavier compositions of cosmic rays have greater lateral spread in the shower, while neutrino-induced events exhibit a small lateral spread. We want to leverage the lateral spread of energy deposits in high-charge cosmic ray showers to reject muon bundles using a Graph Neural Network.
\vspace{-5pt}

\subsection{Rejection Method to Capture Lateral Spread Information}
\begin{figure*}[h!]
    \centering
    \begin{subfigure}[t]{0.5\textwidth}
        \centering
        \includegraphics[width=1\textwidth]{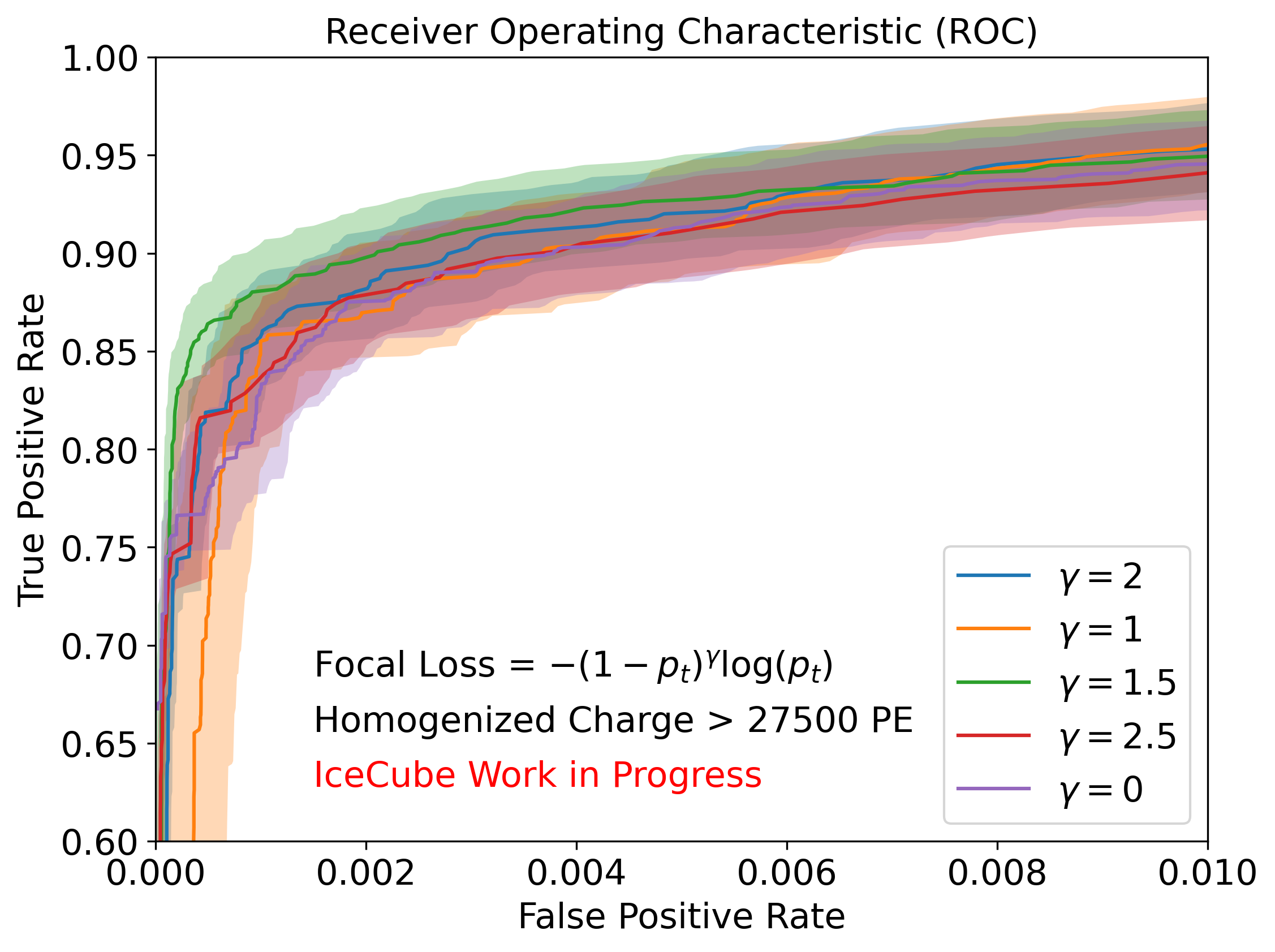}
        \captionsetup{skip=0pt}
        \caption{}
        \label{fig:roc}
    \end{subfigure}%
    ~ 
    \begin{subfigure}[t]{0.5\textwidth}
        \centering
        \includegraphics[width=1\textwidth]{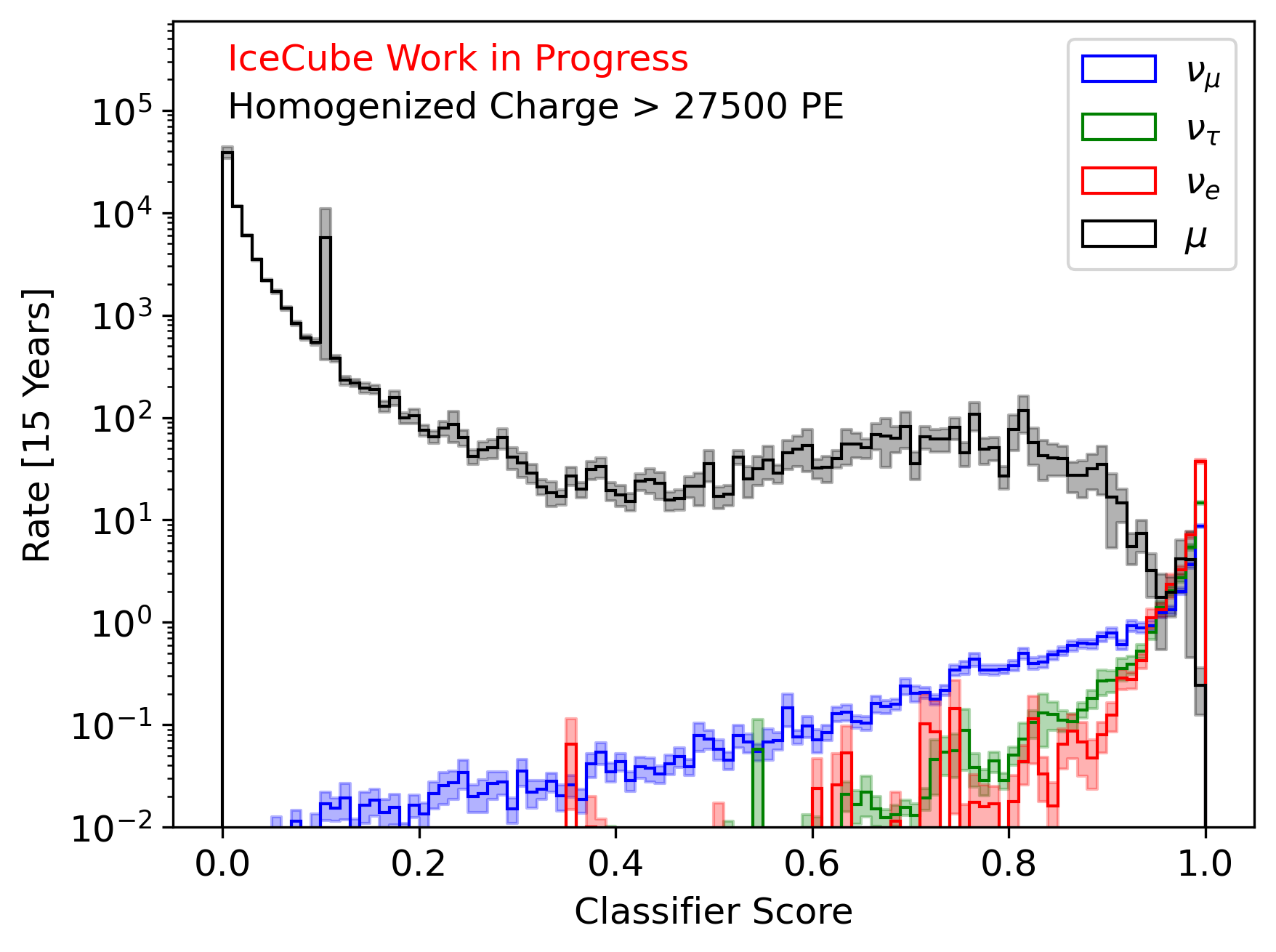}
        \captionsetup{skip=0pt}
        \caption{}
        \label{fig:rates}
    \end{subfigure}
    \captionsetup{skip=0pt}
    \captionsetup{belowskip=-5pt}
    \caption{\textbf{(a)} ROC curve for classifiers trained with different focal loss $\gamma$ parameters described by~\cite{focal}. The focus is on the neutrino signal dominated region of the curve. Astrophysical neutrinos are weighted with the best fit starting track single power law fit and the cosmic rays are weighted with H4a. \textbf{(b)} The score distribution for cosmic rays and neutrinos. Neutrino events have a target of 1, and cosmic ray events have a target of 0. Astrophysical neutrinos are weighted with the best fit starting track single power law measurement~\cite{estes} and the cosmic rays are weighted with GaisserH4a~\cite{h4a}.}
\end{figure*}
We apply the DYNEDGE neural network architecture~\cite{dynedge} using the GraphNeT software package~\cite{graphnetsoftware} to perform a classification task distinguishing neutrinos from muon bundles. As input features per DOM, we include the position in ice (x,y,z), the first hit time, and the total charge. For cosmic ray simulation, we used CORSIKA~\cite{corsika} with the hadronic interaction model SIBYLL 2.3d~\cite{2.3d}. The primary cosmic ray energy distribution followed an $E^{-2}$ power-law generation spectrum. Meanwhile, the neutrino energy generation followed an $E^{-1}$ spectrum. We imposed a minimum charge cut of 27,500 PE on inputs for training. Training weights are applied such that sum of the weights for neutrinos and cosmic ray showers are equal in logarithmic homogenized charge bins. The sum was also made equal across charge bins. Additionally, the neutrino class consisted of 90\% $\nu_{\mu}$ events  and 10\%  $\nu_{\tau}$ events (weighted). The loss function used in training is termed the "Focal" modification to Cross Entropy Loss, which applies a greater loss to greatly misclassified events~\cite{focal}, shown by the equation in \cref{fig:roc}.

\subsection{Rejection Performance}
\begin{figure*}[!b]
    \centering
    \includegraphics[width=1\textwidth]{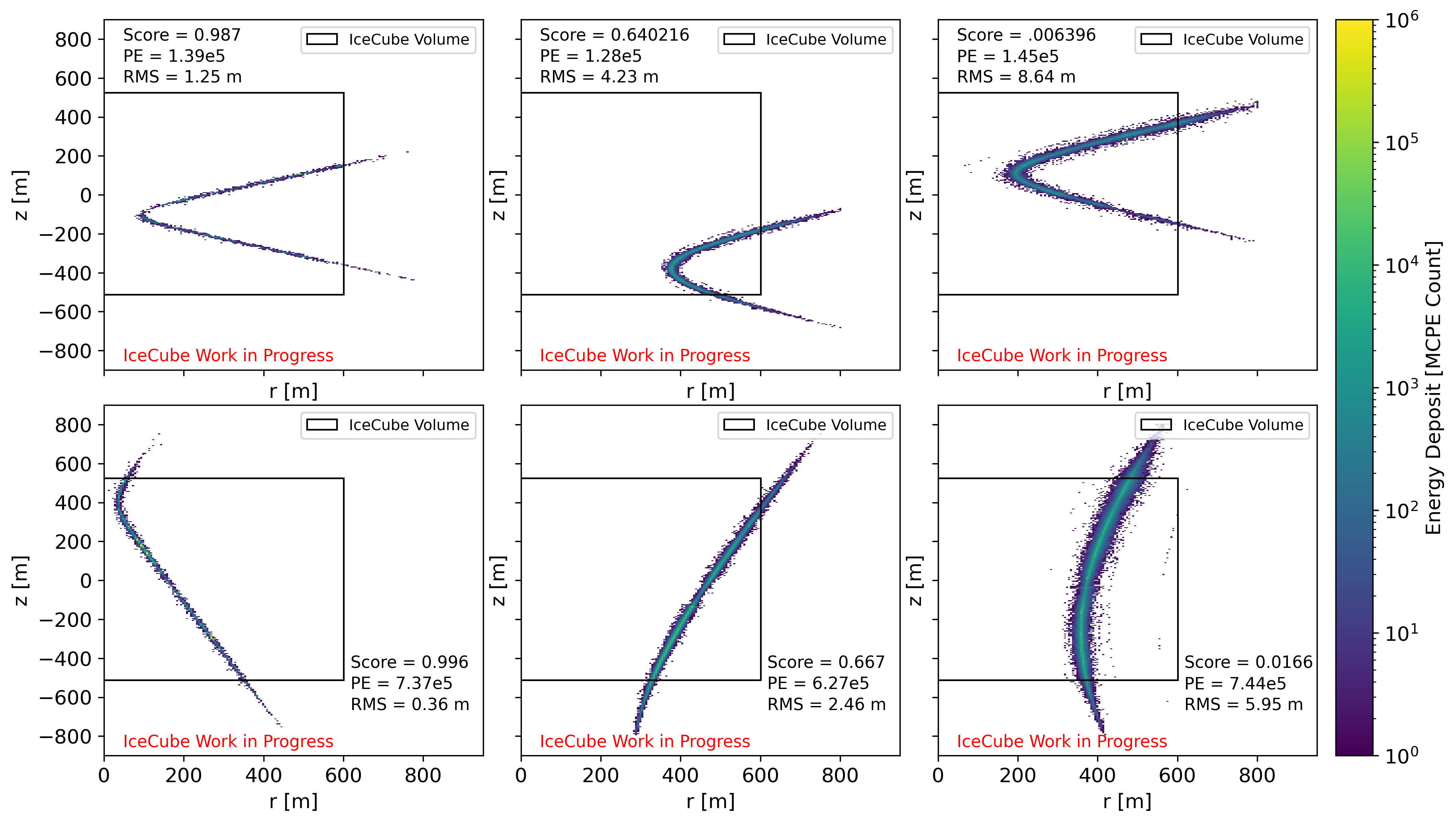}
    \captionsetup{skip=0pt}
    \captionsetup{belowskip=-15pt}
    \caption{Energy deposit locations of cosmic ray showers in IceCube, shown in cylindrical detector coordinates and weighted by the number of photoelectrons detected. The top row shows relatively horizontal cosmic ray showers, while the bottom rows shows more downgoing cosmic ray showers. From left to right, the scores of the showers decrease. High classifier scores are more "neutrino-like" events.}
    \label{fig:single_events}
\end{figure*}
\begin{wrapfigure}{l}{0.475\textwidth}
    \includegraphics[width=.98\linewidth]{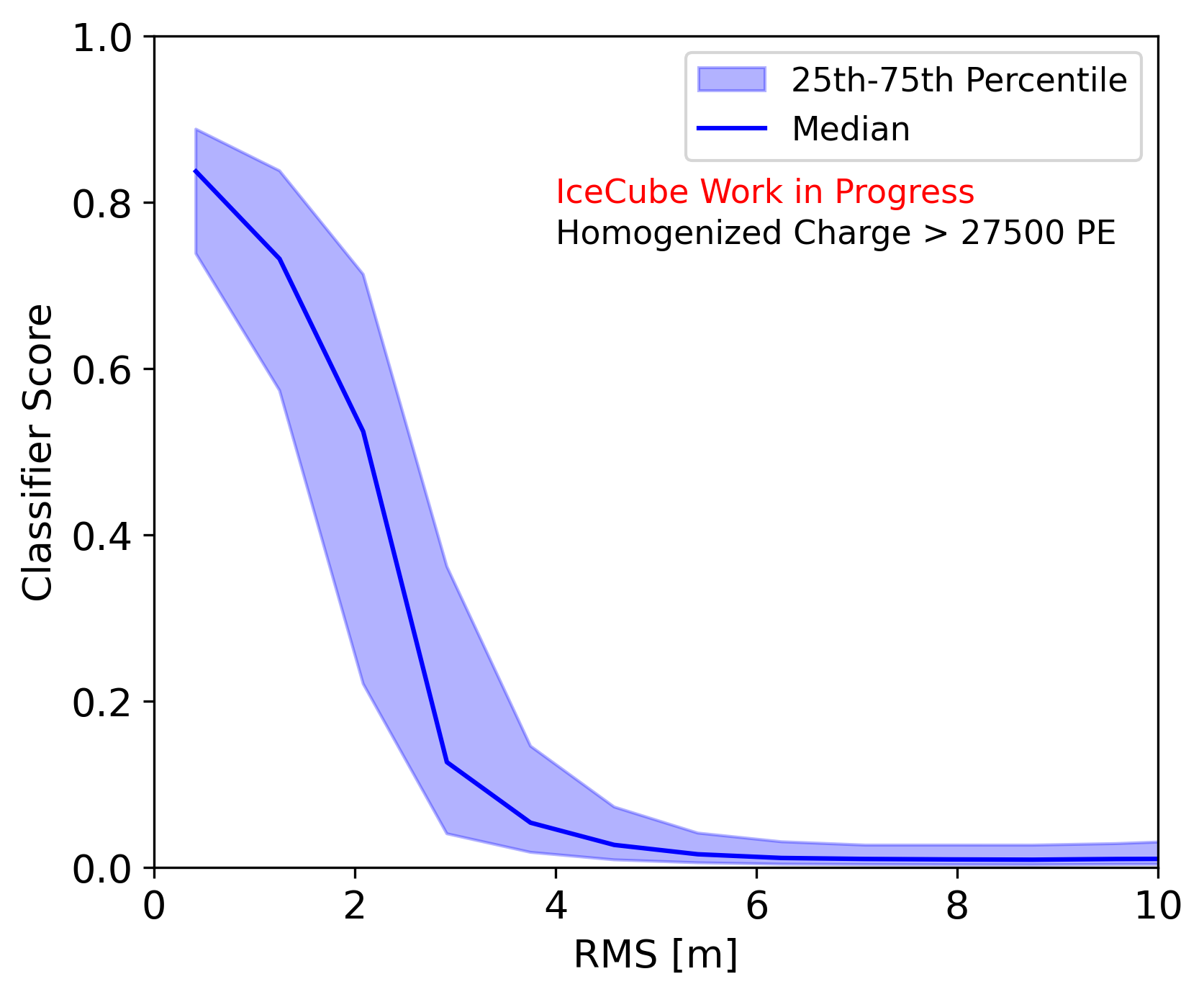}
    \captionsetup{skip=0pt}
    \captionsetup{belowskip=-15pt}
    \caption{Classifier score distribution as a function of the RMS of cosmic ray showers. Percentiles are derived from unweighted Monte Carlo statistics.}
    \label{fig:correlation}
\end{wrapfigure}

This section evaluates the efficiency of the classification. As depicted in \cref{fig:roc}, we show the ROC (Receiver Operating Characteristic) curve for the classifiers trained with different focal loss $\gamma$ parameters~\cite{focal}. Given that a focal loss parameter of $\gamma = 1.5$ shows the best performance, we adopt this classifier for further evaluation. \Cref{fig:rates} depicts the atmospheric muon rate weighted to the GaisserH4a model~\cite{h4a} and the astrophysical neutrino flux is weighted to IceCube's best-fit single power-law flux from the enhanced starting track selection~\cite{estes}. 
\Cref{fig:single_events} depicts how cosmic ray showers of different scores and zenith angles appear in the detector. High-score showers appear thin and more like single-muons, while low-score showers have larger lateral spread and are more distinctly identifiable. The trend is evident in \cref{fig:correlation}, which displays the median, 25th, and 75th percentile classifier scores for cosmic rays based on their RMS values. There is a strong correlation between classification ability and lateral spread of the shower.

\section{Directional Reconstruction}
\subsection{Multitask Transformer based neural network}
\label{sec:intro-mult}

This section describes the Multitask Transformer-based Neural Network (MTNN) used for directional reconstruction in this analysis.

This analysis includes a wide range of event topologies such as through-going tracks, starting tracks, cascades. Therefore, we need a directional reconstruction that makes no assumptions about the topology of the event. 
Previous likelihood based methods, such as SplineMPE, target a specific event topology for reconstruction. 
SplineMPE models a infinite single muon track to reconstruct
direction~\cite{Bradascio_2019}.
The MTNN also simultaneously reconstructs multiple target variables, achieving this by jointly training on multiple tasks. The tasks are split into \textit{supporting} and \textit{final} task(s). By first introducing it to the support task(s), additional information about the final task(s) could be induced. This idea is known as inductive transfer~\cite{Caruana}. 

\begin{figure*}[h!]
    \centering
    \begin{subfigure}[t]{0.6\textwidth}
        \centering
        \includegraphics[width=1\textwidth]{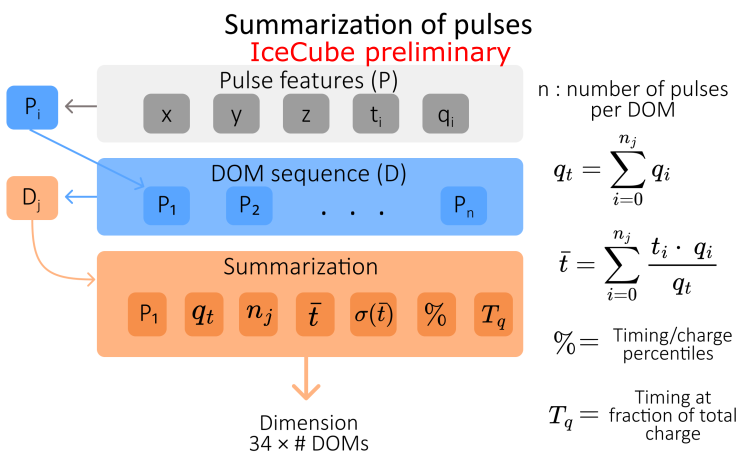}
        \captionsetup{skip=0pt}
        \caption{}
        \label{fig:DOM_summarize}
    \end{subfigure}%
    ~ 
    \begin{subfigure}[t]{0.4\textwidth}
        \centering
        \includegraphics[width=1\textwidth]{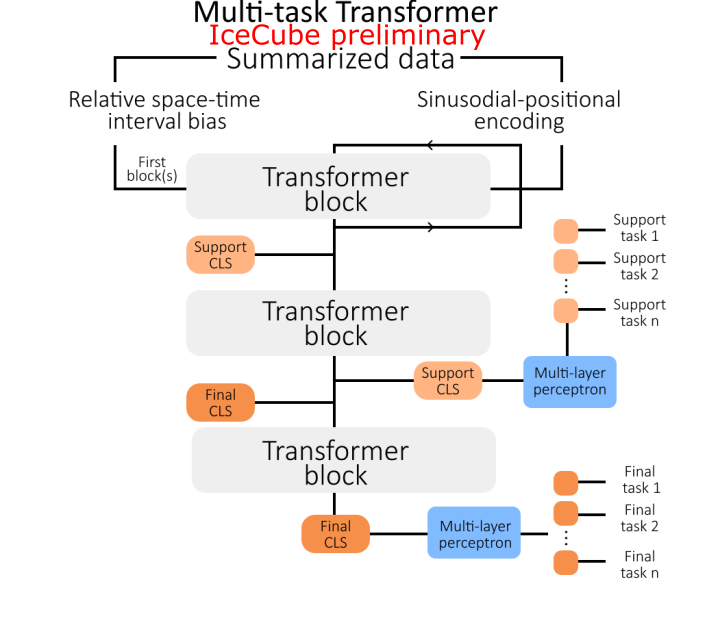}
        \captionsetup{skip=0pt}
        \caption{}
        \label{fig:model_arc}
    \end{subfigure}
    \captionsetup{skip=0pt}
    \captionsetup{belowskip=-10pt}
    \caption{\textbf{(a)} Schematic illustrating how pulses are summarized at the DOM level. \textbf{(b)} Overview of the model architecture, the Transformer blocks as well as the bias and encoding are inspired by the second place solution of the IceCube Kaggle competition~\cite{bukhari2023icecubeneutrinosdeep}.}
\end{figure*}
Another complication of the high energy data is that the number of recorded pulses rises linearly with the energy of the incident particles, necessitating summarization of pulses. By drawing on experience from \cite{Abbasi_2021}, pulse summarization is performed at the DOM level as illustrated in \cref{fig:DOM_summarize}. Pulses were chosen to be summarized per DOM, the total charge, charge weighted mean time, timing/charge percentiles as well as closest times at a given fraction of the total charge are calculated. This summarization significantly reduces the RAM requirements of the GPU.

 The configuration used in this analysis has 2 final tasks and 4 supporting tasks. Some of these tasks are based on the Highest Energy Particle (HEP) in the event, this HEP is the pseudo-particle producing the highest energy deposit in the detector volume extended by 250-meter buffer. The exact definition of the HEP varies depending on the type of event. For through-going tracks, the HEP is the particle producing the track. For through-going bundles the highest energy particle in the bundle determines the track but the energy is the combined energy of the bundle. For starting events the HEP corresponds to the interacting particle at the point of interaction. The following are the truth labels used for training.
\begin{description}[noitemsep,topsep=0.5ex]
    \item[HEP visible length (supporting)]
    The length of the path traveled by the light emitting particle through the detector volume. 
    \item[HEP Trackness (supporting)]
    The amount of energy deposited by track producing particles such as muons, in fractions of the total energy deposit of the HEP.
    \item[HEP Position (supporting)]
    The x,y,z-position of either the interaction vertex of a starting event or the closest approach of a through-going/stopping event.
    \item[Calorimetry energy (supporting)]
    The total amount of energy deposited in the detector by any particle.
    \item[HEP Dir (final)]
    The x,y,z-direction of the HEP.
    \item[HEP energy (final)]
    The energy of the HEP as it becomes visible inside the extended detector volume. For tracks it is defined as the energy of the light producing particle(s) as it enters the volume. For starting events it is defined as the energy of the interaction producing particle at the interaction vertex.
\end{description}
The model architecture can be seen in \cref{fig:model_arc}. The network uses components from the second place solution of the IceCube kaggle competition paper~\cite{bukhari2023icecubeneutrinosdeep} and the overall design is then altered to allow for the Multitask approach. 
The classifier token (CLS) aggregates the information of the full sequence to a fixed length output. By first pulling out the classifier tokens for the support tasks, the final transformer blocks is only used for reconstructing the final task, while inductive bias is gained from the supporting tasks. Variations of the Cauchy loss~\cite{mlotshwa2023cauchylossfunctionrobustness} is used to evaluate each of the tasks separately. 
\subsection{Performance of the directional reconstruction by the network}
\label{sec:perf_dir}

\begin{figure*}[h!]
    \centering
    \begin{subfigure}[t]{0.5\textwidth}
        \centering
        \includegraphics[width=1\textwidth]{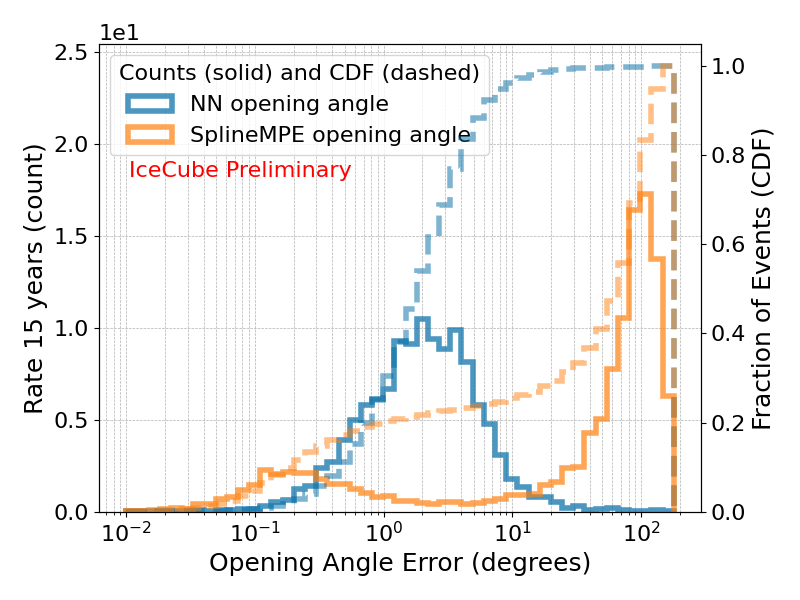}
        \captionsetup{skip=0pt}
        \caption{}
    \end{subfigure}%
    ~ 
    \begin{subfigure}[t]{0.5\textwidth}
        \centering
        \includegraphics[width=1\textwidth]{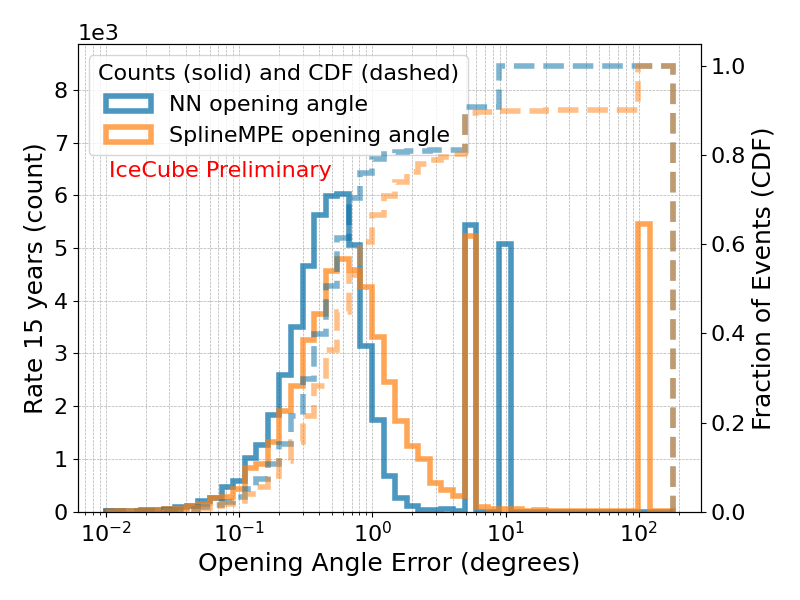}
        \captionsetup{skip=0pt}
        \caption{}
    \end{subfigure}
    \captionsetup{skip=0pt}
    \captionsetup{belowskip=-10pt}
    \caption{ Weighted histogram distribution of the opening angle between reconstructed and true directions. \textbf{(a)} Neutrino events (all flavor) \textbf{(b)} Atmospheric muon background events}
    \label{fig:flavor_hist}
\end{figure*}

As this analysis only makes use of the directional reconstruction of the model, only the performance of this task is evaluated. The model was trained on events exceeding 1,000 PE in homogenized charge. Neutrino events of all flavors as well as downgoing atmospheric muon background events are included in the training. The analysis sample employs a stricter requirement on the homogenized charge of 27,500 PE, such that the sample is compatible with the bundle rejection classifier which is only trained on events above 27,500 PE. Conventional and Prompt neutrinos are weighted according to H3a\_SIBYLL23~\cite{Aartsen_2020}, while astrophysical neutrinos are weighted according to an unbroken powerlaw with $\gamma = -2.52$ and a normalization of $ \Phi_{astro}^0=1.8 \times 10^{-18}~\text{GeV}^{-1}\text{cm}^{-2}\text{s}^{-1}\text{sr}^{-1}$~\cite{globalfit_icrc}. Background muons are weighted using the data-driven global spline fit~\cite{Dembinski:2017zsh}.

\begin{figure*}[h!]
    \centering
    \begin{subfigure}[t]{0.5\textwidth}
        \centering
        \includegraphics[width=1\textwidth]{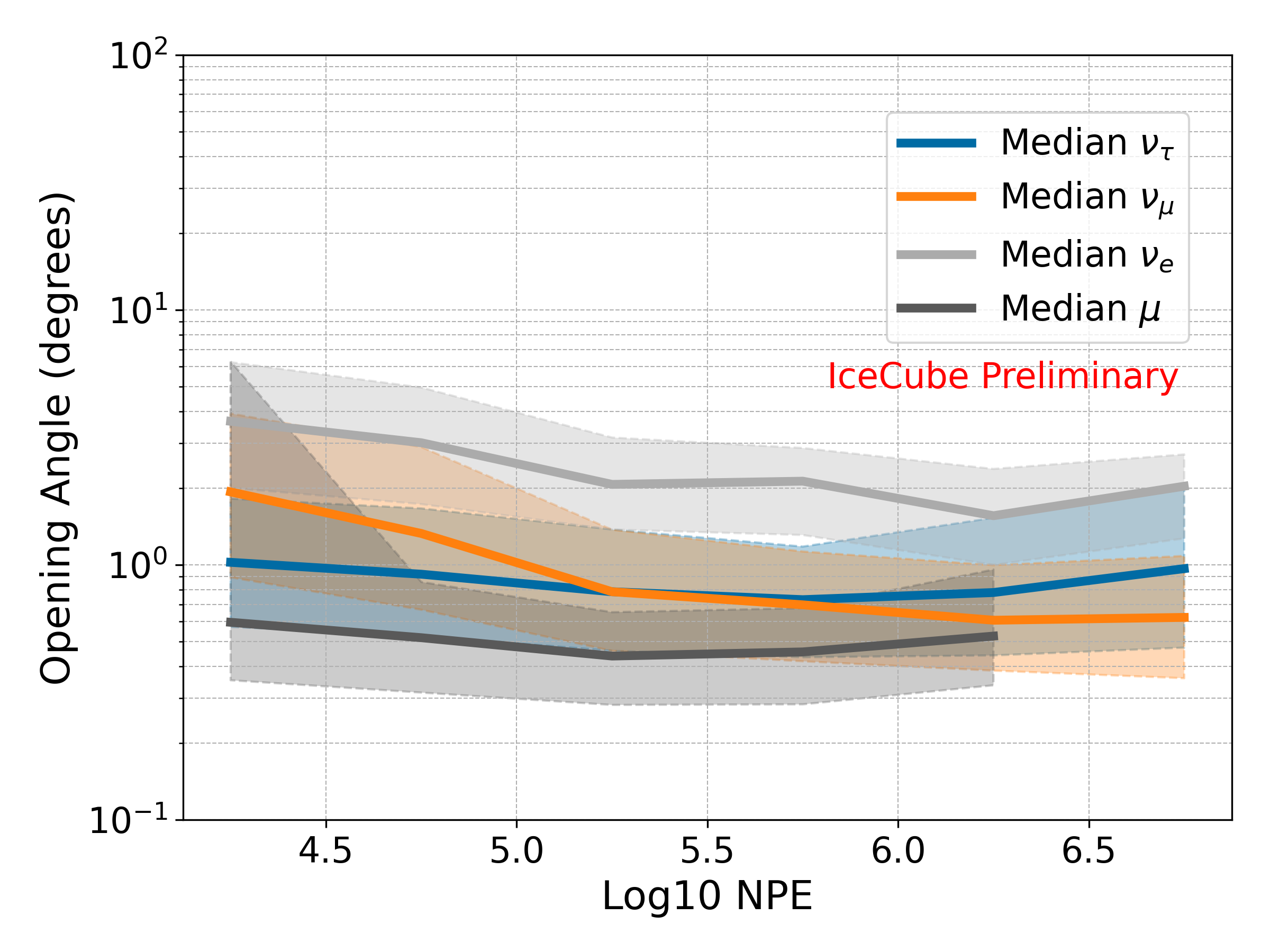}
            \captionsetup{skip=0pt}
        \caption{}
    \end{subfigure}%
    ~ 
    \begin{subfigure}[t]{0.5\textwidth}
        \centering
        \includegraphics[width=1\textwidth]{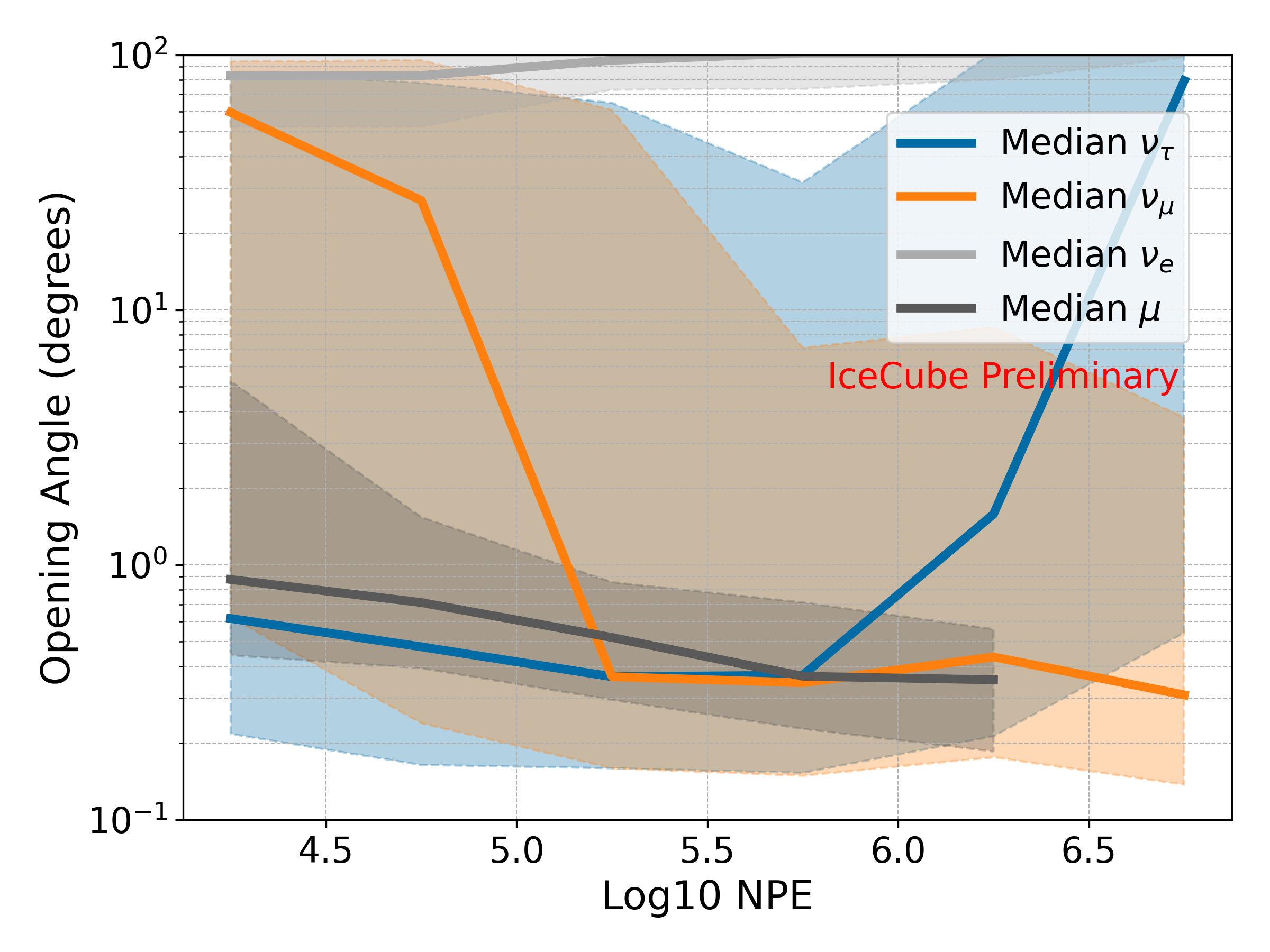}
        \captionsetup{skip=0pt}
        \caption{}
    \end{subfigure}
    \captionsetup{skip=0pt}
    \captionsetup{belowskip=-10pt}
    \caption{\textbf{(a)} MTNN medians and 25/75th percentiles for different neutrino flavors. \textbf{(b)} SplineMPE medians and 25/75th percentiles for different neutrino flavors.}
    \label{fig:flavor_perf}
\end{figure*}

Histogram (a) in \cref{fig:flavor_hist} shows that \textit{SplineMPE} offers a better reconstruction for a small fraction of events, these events are likely to be the ones fitting the prior used to generate the PDF splines. The MTNN starts having more cumulative events at $\sim0.6^\circ$ and at $\sim3^\circ$ has about 3 times the number of events. Histogram (b) shows that the MTNN offers a a better reconstruction over the full range. Fig \ref{fig:flavor_perf} shows that the MTNN offers a more robust reconstruction for this selection of events regardless of the neutrino flavor when compared to SplineMPE. 

\section{Discussion}

\subsection{Application to future Extremely high energy neutrino selections}

Table \ref{tab:rates} Shows the expected 15 year rates of events passing a muon bundle rejection cut of 0.9, below different levels of opening angles to the true direction. The cut is chosen to produce a sample with roughly equal rates of atmospheric muons and astrophysical neutrinos. Specific selections depend on the analysis and others could be chosen. The 15 year rate of atmospheric muons with the same bundle selection is 85 $\pm$ 25. Weights as in in\cref{sec:perf_dir}.

At the loss of well reconstructed events below 0.1$^\circ$, significantly more events below 1$^\circ$ or 5$^\circ$ can be obtained by using the MTNN. A large part of the difference is driven by cascade-like events.

\begin{table}[h]
\captionsetup{belowskip=0pt}
\vspace{-10pt}
\caption{Expected 15 year rates of neutrino events with an opening angle less than the threshold and passing a muon bundle rejection cut of 0.9. The mean rate is the bootstrap mean and the uncertainty is the standard deviation of the bootstrapping. The best performing model at the given threshold has been highlighted.}
\label{tab:rates}
\begin{center}
\vspace{-15pt}
\hspace{-20pt}

\begin{tabular}{|l|cc|cc|}
\hline
 & \multicolumn{2}{|c|}{Mean Rate - NN} & \multicolumn{2}{|c|}{Mean Rate - SplineMPE} \\
\hline
Threshold & All & MC $E_\nu > 10 \text{PeV}$ &  All & MC $E_\nu > 10 \text{PeV}$  \\
< 0.1 $^\circ$ & 0.28 $\pm$ 0.06 & 0.039 $\pm$ 0.004 & \textbf{2.08 $\pm$ 0.18} &\textbf{ 0.45 $\pm$ 0.023} \\
< 1 $^\circ$ & \textbf{20.8 $\pm$ 0.9} & \textbf{2.3 $\pm$ 0.04} &  12.5 $\pm$ 0.7 & 2.09 $\pm$ 0.04 \\
< 5 $^\circ$  & \textbf{88.6 $\pm$ 2.1} & \textbf{5.39 $\pm$ 0.07} & 15.7 $\pm$ 0.8 & 2.62 $\pm$ 0.05 \\
$\geq$ 5$^\circ$  & \textbf{18.4 $\pm$ 0.8} & \textbf{0.759 $\pm$ 0.025} & 91.2 $\pm$ 2.1 & 3.51 $\pm$ 0.06 \\
\hline
\end{tabular}
\end{center}
\vspace{-1cm}
\end{table}

\section{Conclusion}
In this study, we applied machine learning techniques to reject the background of atmospheric muons from cosmic ray showers and to improve the directional reconstruction of EHE neutrinos. The muon bundle rejection method demonstrated the ability to classify cosmic ray showers, correlating with the RMS of laterally spread energy deposits in the detector from muon bundles. Additionally, the MTNN model showed accurate reconstruction performance across all event topologies compared to SplineMPE. Utilizing these techniques in conjunction with the background rejection methods employed in prior EHE neutrino searches, including the muon overburden and longitudinal deposits (stochastic loss profile), we aim to suppress additional background in the downgoing region to detect EHE neutrinos and improve measurements of diffuse fluxes at energies greater than 10 PeV.

\bibliographystyle{ICRC}
\setlength{\bibsep}{0pt}
\small{\bibliography{refs}}

\clearpage

\section*{Full Author List: IceCube Collaboration}

\scriptsize
\noindent
R. Abbasi$^{16}$,
M. Ackermann$^{63}$,
J. Adams$^{17}$,
S. K. Agarwalla$^{39,\: {\rm a}}$,
J. A. Aguilar$^{10}$,
M. Ahlers$^{21}$,
J.M. Alameddine$^{22}$,
S. Ali$^{35}$,
N. M. Amin$^{43}$,
K. Andeen$^{41}$
% ...
\\
\\
% ...
$^{10}$ Universit{\'e} Libre de Bruxelles, Science Faculty CP230, B-1050 Brussels, Belgium \\
% ...
$^{16}$ Department of Physics, Loyola University Chicago, Chicago, IL 60660, USA \\
$^{17}$ Dept. of Physics and Astronomy, University of Canterbury, Private Bag 4800, Christchurch, New Zealand \\
% ...
$^{21}$ Niels Bohr Institute, University of Copenhagen, DK-2100 Copenhagen, Denmark \\
$^{22}$ Dept. of Physics, TU Dortmund University, D-44221 Dortmund, Germany \\
% ...
$^{35}$ Dept. of Physics and Astronomy, University of Kansas, Lawrence, KS 66045, USA \\
% ...
$^{39}$ Dept. of Physics and Wisconsin IceCube Particle Astrophysics Center, University of Wisconsin{\textemdash}Madison, Madison, WI 53706, USA \\
% ...
$^{41}$ Department of Physics, Marquette University, Milwaukee, WI 53201, USA \\
% ...
$^{43}$ Bartol Research Institute and Dept. of Physics and Astronomy, University of Delaware, Newark, DE 19716, USA \\
% ...
$^{63}$ Deutsches Elektronen-Synchrotron DESY, Platanenallee 6, D-15738 Zeuthen, Germany \\
$^{\rm a}$ also at Institute of Physics, Sachivalaya Marg, Sainik School Post, Bhubaneswar 751005, India \\
% ...

\subsection*{Acknowledgments}

\noindent
The authors gratefully acknowledge the support from the following agencies and institutions:
USA {\textendash} U.S. National Science Foundation-Office of Polar Programs,
U.S. National Science Foundation-Physics Division,
U.S. National Science Foundation-EPSCoR,
U.S. National Science Foundation-Office of Advanced Cyberinfrastructure,
Wisconsin Alumni Research Foundation,
Center for High Throughput Computing (CHTC) at the University of Wisconsin{\textendash}Madison,
Open Science Grid (OSG),
Partnership to Advance Throughput Computing (PATh),
Advanced Cyberinfrastructure Coordination Ecosystem: Services {\&} Support (ACCESS),
Frontera and Ranch computing project at the Texas Advanced Computing Center,
U.S. Department of Energy-National Energy Research Scientific Computing Center,
Particle astrophysics research computing center at the University of Maryland,
Institute for Cyber-Enabled Research at Michigan State University,
Astroparticle physics computational facility at Marquette University,
NVIDIA Corporation,
and Google Cloud Platform;
Belgium {\textendash} Funds for Scientific Research (FRS-FNRS and FWO),
FWO Odysseus and Big Science programmes,
and Belgian Federal Science Policy Office (Belspo);
Germany {\textendash} Bundesministerium f{\"u}r Forschung, Technologie und Raumfahrt (BMFTR),
Deutsche Forschungsgemeinschaft (DFG),
Helmholtz Alliance for Astroparticle Physics (HAP),
Initiative and Networking Fund of the Helmholtz Association,
Deutsches Elektronen Synchrotron (DESY),
and High Performance Computing cluster of the RWTH Aachen;
Sweden {\textendash} Swedish Research Council,
Swedish Polar Research Secretariat,
Swedish National Infrastructure for Computing (SNIC),
and Knut and Alice Wallenberg Foundation;
European Union {\textendash} EGI Advanced Computing for research;
Australia {\textendash} Australian Research Council;
Canada {\textendash} Natural Sciences and Engineering Research Council of Canada,
Calcul Qu{\'e}bec, Compute Ontario, Canada Foundation for Innovation, WestGrid, and Digital Research Alliance of Canada;
Denmark {\textendash} Villum Fonden, Carlsberg Foundation, and European Commission;
New Zealand {\textendash} Marsden Fund;
Japan {\textendash} Japan Society for Promotion of Science (JSPS)
and Institute for Global Prominent Research (IGPR) of Chiba University;
Korea {\textendash} National Research Foundation of Korea (NRF);
Switzerland {\textendash} Swiss National Science Foundation (SNSF).

\end{document}